\documentclass{llncs}
\usepackage{latexsym}
\usepackage{amssymb}
\usepackage{amsmath}
\usepackage{graphicx}
\usepackage{multirow}
\usepackage[nice]{nicefrac}
\usepackage{enumerate}
\usepackage{verbatim} 
\usepackage{color}
\usepackage{hyperref}
\usepackage{listings}
\usepackage{float}
\usepackage[titletoc,toc,title]{appendix}

\newcommand{\nop}[1]{}

\newcommand{\nhalf}{$\nicefrac{1}{2}$}

\newcommand{\MFL}{{\sc Mfl}}
\newcommand{\EMFL}{{\sc Emfl}}
\newcommand{\PFL}{{\sc Pfl}}
\newcommand{\GFL}{{\sc Gfl}}
\newcommand{\tool}{{\sc TTP}}

\floatstyle{plain} % optionally change the style of the new float
\newfloat{datafile}{H}{myc}
\title{\tool : Tool for Tumor Progression}
\titlerunning{\tool}
\author{Johannes G. Reiter\inst{1}, Ivana Bozic\inst{2,3}, Krishnendu Chatterjee\inst{1}, Martin A. Nowak\inst{2,3,4}}
\institute{IST Austria (Institute of Science and Technology Austria), Klosterneuburg, Austria
	\and Program for Evolutionary Dynamics, Harvard University, Cambridge, USA
	\and Department of Mathematics, Harvard University, Cambridge, USA 
	\and Department of Organismic and Evolutionary Biology, Harvard University, Cambridge, USA}
\date{\today}

\begin{document}
 
\maketitle

\begin{abstract}
In this work we present a flexible tool for tumor progression, which simulates the evolutionary dynamics of cancer. Tumor progression implements a multi-type branching process where the key parameters are the fitness landscape, the mutation rate, and the average time of cell division. The fitness of a cancer cell depends on the mutations it has accumulated. The input to our tool could be any fitness landscape, mutation rate, and cell division time, and the tool produces the growth dynamics and all relevant statistics.
\end{abstract}
\vspace{-2.0em}
\section{Introduction}
\vspace{-0.5em}
Cancer is a genetic disease which is driven by the somatic evolution of cells~\cite{vogelstein2004},
where driver mutations for cancer increase the reproductive rate of cells through different mechanisms, 
e.g. evading growth suppressors, sustaining proliferative signaling, or resisting cell death~\cite{Hanahan2011a}.
Tumors are initiated by some genetic event which increases the reproductive rate of previously 
normal cells.
The evolution of cancer (malignant tumor) is a multi-step process where cells need to 
receive several mutations subsequently~\cite{Jones2008evolution}. 
This phase of tumor progression is characterized by the uncontrolled growth of 
cells~\cite{Hanahan2011a,Nowak2006}.
The requirement to accumulate multiple mutations over time explains the 
increased risk of cancer with age.

There are several mathematical models to explain tumor progression and the age incidence of 
cancer~\cite{komarova2003mutation,iwasa2004tunnels,nowak2004inactivation}.
The models have also provided quantitative insights in the evolution 
of resistance to cancer therapy~\cite{Diaz2012}.
The models for tumor progression are multi-type branching processes which represent 
an exponentially growing heterogeneous population of cells, where 
the key parameters for the process are: 
(i)~the fitness landscape of the cells (which determine the reproductive rate), 
(ii)~the mutation rate (which determines the accumulation of driver mutations), 
and (iii)~the average cell division time (or the generation time for new cells).
The fitness landscapes allow the analysis of the effects of interdependent 
(driver) mutations on the evolution of cancer \cite{Bozic2010}.

In this work, we present a very flexible tool (namely, \tool, tool for tumor progression) 
to study the dynamics of tumor progression.
The input to our tool could be any fitness landscape, mutation rate, and 
cell division time, and the tool generates the growth dynamics and all relevant 
statistics (such as the expected tumor detection time, or the expected
appearance time of surviving mutants, etc).
Our stochastic computer simulation is an efficient simulation of a multi-type
branching process under all possible fitness landscapes, driver mutation rates, and cell division times.

Our tool provides a quantitative framework to study the dynamics of tumor progression 
in different stages of tumor growth. 
Currently, the data to understand the effects of complex fitness landscapes can only
be obtained from patients or animals suffering the disease.
With our tool, playing with the parameters, once the real-world data is reproduced,
the computer simulations can provide many simulation examples that would aid 
to understand these complex effects.
Moreover, once the correct mathematical models for specific types of cancer are
identified where the simulations match the real-world data, 
verification tools for probabilistic systems can be used to further analyze
and understand the tumor progression process (such an approach has been followed in~\cite{sadot2008toward} for the verification of biological models).
In this direction, results of specific fitness landscapes of our tool have already 
been used in a biological application paper~\cite{Reiter2012onebit}.
While we present our tool for the discrete-time process (which provides a 
good approximation of the continuous-time process), results of our tool for 
the special case of a uniform fitness landscape in the continuous-time process 
have also been shown to have excellent agreement with the 
real-world data for the time to treatment failure for colorectal cancer~\cite{Diaz2012}.

\section{Model} \label{sec:model}
Tumor progression is modeled as a discrete-time branching process (Galton-Watson process \cite{Haccou2005}).
At each time step, a cell can either divide or die.
The phenotype $i$ of a cancerous cell determines its division probability $b_i$ and is encoded as a bit string of length four (i.e.~$\{0,1 \}^4$).
The death probability $d_i$ follows from $b_i$ as $d_i = 1 - b_i$.
If a cell divides, one of the two daughter cells can receive an 
additional mutation (i.e., a bit flips from wildtype $0$ to the mutated type $1$) 
with probability $u$ in one of the wildtype positions (e.g., cells of phenotype $1010$ can receive an additional mutation only at positions two and four; cells of phenotype $1111$ can not receive any additional mutations).
The branching process is initiated by a single cell of phenotype~$i=0000$ (resident cell).
The resident cells are wildtype at all four positions and have a strictly positive growth rate (i.e., $b_{0000}-d_{0000} > 0$). %%%which results in a super-critical branching process.

\smallskip\noindent{\bf Fitness landscapes.}
Our tool provides two predefined fitness landscapes for driver mutations in tumor progression: 
(1)~Multiplicative Fitness Landscape (\MFL) and (2)~Path Fitness Landscape (\PFL).
Additionally, the user can also define its own general fitness landscape (\GFL).
A fitness landscape defines the birth probability $b_i$ for all possible phenotypes $i$.
Following the convention of the standard modeling approaches, we let $b_{0000} = \nhalf(1 + s_0)$ be the birth probability of the resident cells (i.e., cells of phenotype~$0000$) \cite{Bozic2010,Reiter2012onebit}. 
The growth coefficient $s_j$ indicates the selective advantage provided by an additional mutation at position~$j$ in the phenotype.

\smallskip\noindent
{\em Multiplicative fitness landscape.}
In the \MFL\ a mutation at position $j$ of the phenotype $i$ of a cell results in a multiplication of its birth probability by $(1 + s_j)$.
Specifically, the birth probability $b_i$ of a cell with phenotype $i$ is given by
\[
b_i = \frac{1}{2} (1+s_0) \prod_{j=1}^{4} {(1+\widehat{s}_j)};
\]
where $\widehat{s}_j=0$ if the $j$-th position of $i$ is~0; otherwise
$\widehat{s}_j=s_j$.
Hence, each additional mutation can be weighted differently and provides a predefined 
effect $(1+s_1)$, $(1+s_2)$, $(1+s_3)$, or $(1+s_4)$ on the birth probability of a cell.
Additional mutations can also be costly (or neutral) which can be modeled by a negative $s_j$ (or $s_j=0$).
If $s_0=s_1=s_2=s_3=s_4$ the fitness landscape reduces to the model studied by Bozic et al. \cite{Bozic2010},
which we call \EMFL\ (Equal Multiplicative Fitness Landscape) and is also predefined in our tool.
%In Table \ref{tab:mfl} we present the details of the \MFL .

\smallskip\noindent
{\em Path fitness landscape.}
The \PFL\ defines a certain path on which additional mutations need to occur to increase the birth probability of a cell.
The predefined path can be $0000 \rightarrow 1000 \rightarrow 1100 \rightarrow 1110 \rightarrow 1111$, and
again the growth coefficients $s_j$ determine the multiplicative effect of the new mutation on the birth probability (see appendix for more details).
Mutations not on this path are deleterious for the growth rate of a cell and its birth probability is set to~\nhalf$(1 - v)$.
The parameter~$v$ ($0 \leq v \leq 1$) specifies the disadvantage for cells of all phenotypes which do not belong to the given path.  

\smallskip\noindent{\em General fitness landscapes.} Our tool allows to input 
any fitness landscape as follows: for $b_i$ for $i\in \{0,1\}^4$, our tool
can take as input the value of $b_i$. In this way, any fitness landscape can be
a parameter to the tool.

\smallskip\noindent{\bf Density limitation.}
In some situations, a tumor needs to overcome current geometric or metabolic constraints 
(e.g. when the tumor needs to develop blood vessels to provide enough oxygen and nutrients for further growth \cite{Kerbel2000,Reiter2012onebit}).
Such growth limitations are modeled by a density limit (carrying capacity) for various phenotypes.
Hence, the cells of a phenotype~$i$ grow first exponentially but eventually reach a steady state around a given carrying capacity $K_i$.
Only cells with another phenotype (additional mutation) can overcome the density limit.
Logistic growth is modeled with variable growth coefficients $\widetilde{s_j} = s_j (1 - X_i / K_i)$ where $X_i$ is the current number of cells of phenotype~$i$ in the tumor.
In this model, initially $\widetilde{s_j} \approx s_j$ ($X_i \ll K_i$), however, if $X_i$ is on the order of $K_i$, $\widetilde{s_j}$ becomes approximately zero (details are given in the appendix).

\section{Tool Implementation \& Experimental Results}

Our tool provides an efficient implementation of a very general tumor progression model.
Essentially, the tool implements the above defined branching processes to simulate the dynamics of tumor growth and to obtain statistics about the expected tumor detection time and the appearance of additional driver mutations during different stages of disease progression.
\tool\ can be downloaded from here:~\url{http://pub.ist.ac.at/ttp}.

For an efficient processing of the discrete-time branching process, the stochastic simulation samples from a multinomial distribution for each phenotype at each time step \cite{Bozic2010,Reiter2012onebit}.
The sample returns the number of cells which divided with and without mutation and the number of cells which died in the current generation (see the appendix for details).
From the samples for each phenotype the program calculates the phenotype distribution in the next generation.
Hence, the program needs to store only the number of cells of each phenotype during the simulation.
This efficient implementation of the branching process allows the tool to simulate many patients within a second and to obtain very good statistical results in a reasonable time frame.

\begin{figure}[h]
	\centering
	\includegraphics[width=0.49\textwidth]{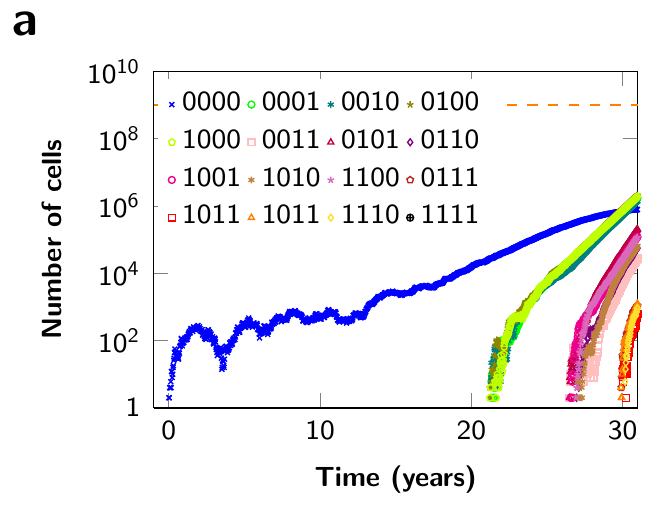}
	\includegraphics[width=0.49\textwidth]{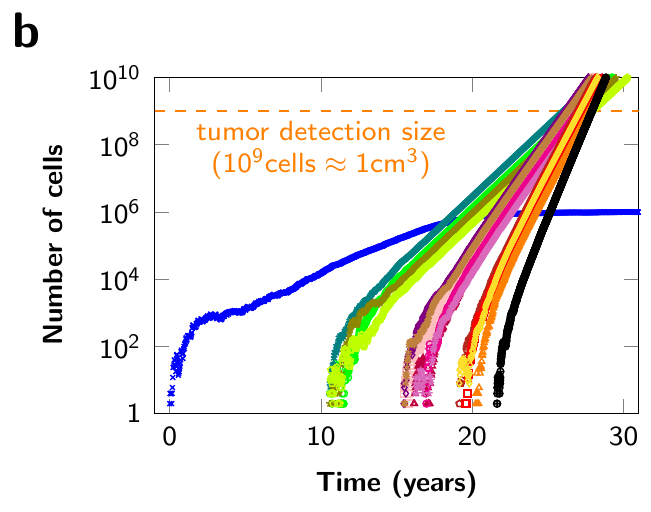}
	\includegraphics[width=0.49\textwidth]{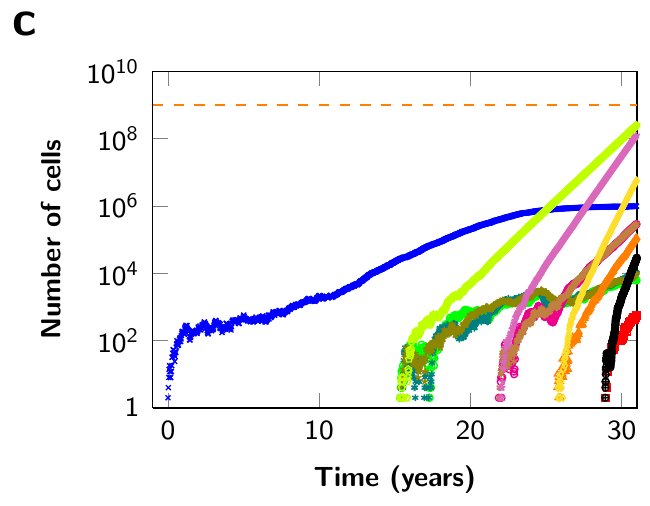}
	\includegraphics[width=0.49\textwidth]{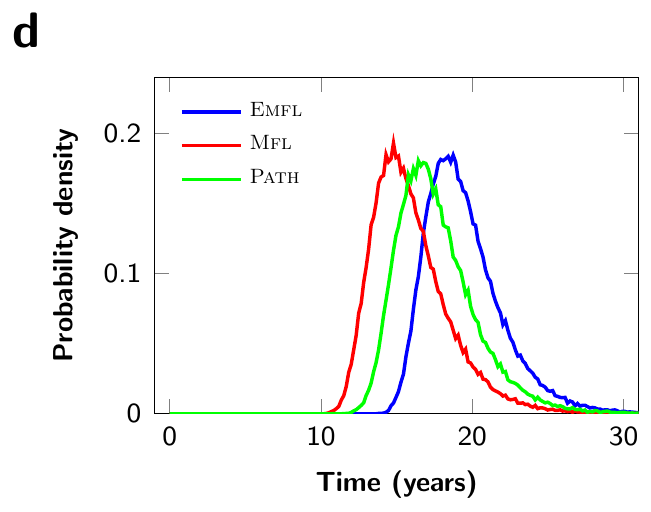}
	\caption{
	Experimental results illustrating the variability of tumor progression.
	In panels a and b we show examples for two particular simulation runs 
	where the cells grow according to the \EMFL\ and resident cells (blue) are constrained by a carrying capacity of $K_{0000}=10^6$.
	In panel c the cells grow according to the \PFL.
	In panel d we show statistical results for the probability density of tumor detection
	when cells grow according to different fitness landscapes.
	Parameter values: growth coefficients $s_0=0.004$, $s_1=0.006$, $s_2=0.008$ , $s_3=0.01$, $s_4=0.012$, and $v=0.01$, mutation rate $u=10^{-6}$, cell division time $T=3$~days, tumor detection size $10^9$ cells.
	}
	\label{fig:examples}
\end{figure}

\smallskip\noindent
\textbf{Modes.} 
The tool can run in the following two modes: individual or statistics.
In the \textit{individual mode} the tool produces the growth dynamics of one tumor in a patient (see panels a, b, and c in Fig. \ref{fig:examples}). 
Furthermore, both the growth dynamics and the phenotype distribution of the tumor are depicted graphically.
In the \textit{statistics mode} the tool produces the probability distribution for the detection time of the tumor (see panel d in Fig. \ref{fig:examples}) both graphically and quantitatively.
Additionally, the tool calculates for all phenotypes the appearance times of the first surviving lineage, the existence probability, and the average number of cells at detection time.

\smallskip\noindent
\textbf{Features.}
%Our tool efficiently generates the probability density of tumor detection for any input parameters and the expected appearance times of \jgr{improve}
\tool\ provides an intuitive graphical user interface to enter the parameters of the model and shows plots of the dynamics during tumor progression, the phenotype distribution, or the probability density of tumor detection.
These plots can also be saved as files in various image formats.
Furthermore, the tool can create data files (tab-separated values) of the tumor growth history and the probability distribution of tumor detection for any set of input parameters (details on the format are given in the appendix). 

\smallskip\noindent
\textbf{Input parameters.}
In both modes, the tool takes the following input parameters: (i) growth coefficients $s_0$, $s_1$, $s_2$, $s_3$, and $s_4$ (and $v$ in the case of \PFL ), (ii) mutation rate $u$, (iii) cell generation time $T$, (iv) fitness landscape (\MFL , \PFL , \EMFL , or \GFL\ with the birth probability for each phenotype), and optional (v) density limits for some phenotypes.
In the individual mode, additionally, the user needs to provide the number of generations which have to be simulated.
In the statistics mode, the additional parameters are: the tumor detection size and the number of patients (tumors which survive the initial stochastic fluctuations) which have to be simulated.

\smallskip\noindent
\textbf{Experimental results.}
In panels a, b, and c of Fig. \ref{fig:examples} we show examples of the growth dynamics during tumor progression.
Although we used exactly the same parameters in panels a and b, we observe that the time from tumor initiation until detection can be very different.
%However, 30 years after tumor initiation, the tumor consists of roughly $10^6$ cells in panel a whereas in panel b the tumor consists of approximately $10^{11}$ cells.
In panel d we show the probability density of tumor detection under various fitness landscapes.
Further experimental results are given in the appendix.

\smallskip\noindent
\textbf{Case studies.}
Several results of these models have shown excellent agreement with different aspects of real-world data.
In \cite{Bozic2010}, results for the expected tumor size at detection time using a \EMFL\ fit the reported polyp sizes of the patients very well.
Similarly, using a continuous-time branching process and a uniform fitness landscape, results for the expected time to the relapse of a tumor after start of treatment agree thoroughly with the observed times in patients \cite{Diaz2012}.

\smallskip\noindent
\textbf{Future work.}
In some ongoing work, we also investigate mathematical models for tumor dynamics occurring during 
cancer treatment modeled by a continuous-time branching process.
Thus an interesting extension of our tool would be to model treatment as well. 
Another interesting direction is to model the seeding of metastasis during 
tumor progression and hence simulate a ``full'' patient rather than the 
primary tumor alone.
Once faithful models of the evolution of cancer have been identified, 
verification tools such as PRISM~\cite{PRISM} and theoretical results such 
as~\cite{ESY12} might contribute to the understanding of these processes.

\smallskip\noindent
\textbf{Acknowledgments.}
This work is supported by the ERC Start grant (279307: Graph Games), the FWF NFN Grant (No S11407-N23, Rise), the FWF Grant (No P 23499-N23), a Microsoft Faculty Fellow Award, the Foundational Questions in Evolutionary Biology initiative of the John Templeton Foundation, and the NSF/NIH joint program in mathematical biology (NIH grant R01GM078986).

%\bibliographystyle{splncs}
%\bibliography{cancer}

\clearpage
\appendix

\section{Appendix: Details of the Tool}
\renewcommand\thetable{\thesection\arabic{table}}
\renewcommand\thefigure{\thesection\arabic{figure}}
\renewcommand\lstlistingname{Listing}
\renewcommand\thelstlisting{\thesection\arabic{lstlisting}}

\tool\ is available for download at \url{http://pub.ist.ac.at/ttp}.
The tool is implemented in Java and runs on all operating systems which run a Java Virtual Machine (JVM) of version 1.7 or above (\texttt{\href{http://www.java.com/getjava}{www.java.com/getjava}}).
All the necessary libraries are included in the jar-file.

\subsection{Tool Features}
Our tool supports various features in two running modes. %(i) individual mode and (ii) statistics mode.

\smallskip
%\noindent
%\textbf{Individual mode.} 
In the individual mode, \tool\ simulates the tumor growth dynamics for a given number of generations.
Plots of the growth dynamics over time and the current phenotype distribution are produced simultaneously.
(Both plots can be saved in a PNG-file or SVG-file.) 
The full growth history for all cell types can also be stored in a data-file (format is described in section \ref{app:impl}).

\smallskip
%\noindent
%\textbf{Statistics mode.} 
In the statistics mode, \tool\ simulates the given number of patients with the same parameters and simultaneously shows the probability density of tumor detection for a given detection size ($10^9$ cells correspond to a tumor volume of approximately $1\mathrm{cm^3}$).
The average tumor detection time and the average fraction of resident cells at detection are also shown during the simulations.
After all patients have been simulated, the existence probability at detection, the average number of cells, and the average appearance year of the first surviving cell for all phenotypes are calculated and shown in a new window.
In addition the tool shows the number of detected and died tumors per year in a separate window.
All these data is stored in a data-file (format is described in section \ref{app:impl}).

\subsection{Installation and Implementation Details} \label{app:impl}

\tool\ is written in Java and makes use of several other libraries.
The tool requires the Java Runtime Environment (JRE) of version 1.7 or above.
To start \tool\ double-click on \texttt{ttp.jar} or on the command line type: \texttt{java -jar ttp.jar}. (Make sure that you have the permission to execute \texttt{ttp.jar}. On Mac OS invoking the tool from the command line can overcome the security restrictions.)

\smallskip
The tool is composed of the following components: the model implementation, the statistics thread, the graphical user interface, and the plot generator.

\smallskip\noindent
\textbf{Model implementation.} 
The core component of the tool is the efficient implementation of the discrete-time branching process.
Following Bozic et al. \cite{Bozic2010}, the number of cells $X_i$ of phenotype $i$ in the next generation $(t+1)$ is calculated by sampling from the multinomial distribution
\begin{equation}
\mathrm{Prob} [(Y_1,Y_2,Y_3) = (y_1,y_2,y_3)] = \frac{X_i(t) !}{y_1 ! y_2 ! y_3 !} [b_i (1-u)^{y_1} d_i^{y_2} (b_i u)^{y_3}] 
\end{equation}
where $y_1 + y_2 + y_3 = X_i (t)$, $y_3 = \sum_k{M_{ik}}$, and 
\begin{equation}
X_i (t+1) = X_i (t) + Y_1 - Y_2 + \sum_k{M_{ki}} \ .
\end{equation}
The number of cells which give birth to an identical daughter cell is denoted by $Y_1$,
the number of cells which die is denoted by $Y_2$.
The number of cells which divide with an additional mutation is given by $Y_3$ and the number of cells mutated from phenotype $k$ to $i$ is given by $M_{ki}$.
In general, one can define a mutation matrix to encode the probabilities $m_{ki}$ that a cell of phenotype $k$ mutates to a cell of phenotype $i$.
In our case this matrix is defined by the sequential accumulation of mutations.
A cell of some phenotype can receive an additional mutation only on the positions in its bit-string encoding which are wildtype (i.e. only bit flips from 0 to 1 are allowed).
Mutations on all allowed positions are equally likely.
Back mutations are not considered.

\smallskip\noindent{\em Fitness landscapes.}
Our tool supports four fitness landscapes for additional driver and passenger mutations: (i) \MFL , (ii) \EMFL , (iii) \PFL , and (iv) \GFL . 
In principal, driver mutations increase the birth rate of a cell whereas passenger mutations have no effect on the cell's birth rate \cite{Bozic2010}.
In the tables~\ref{tbl:mfl} and~\ref{tbl:pfl} we present the complete definition of the \MFL\ and the \PFL, respectively.
(The definition of \EMFL\ and \GFL\ have been given in Section \ref{sec:model}.)

\begin{table}[h]
\small
\centering
\caption{Multiplicative Fitness Landscape.}
%\centering
	\begin{tabular}{c c c}
\hspace{0.2cm} Additional mutations \hspace{0.2cm} & \multirow{1}{*}{Phenotype} & \multirow{1}{*}{Birth probability} \\
\hline	
0 & 0000 & $\nicefrac{1}{2}(1+s_0)$ \\
\hline
\multirow{4}{*}{1} & 1000 & $\nicefrac{1}{2}(1+s_0)(1+s_1)$ \\
& 0100 & $\nicefrac{1}{2}(1+s_0)(1+s_2)$ \\
& 0010 & $\nicefrac{1}{2}(1+s_0)(1+s_3)$ \\
& 0001 & $\nicefrac{1}{2}(1+s_0)(1+s_4)$ \\
\hline

\multirow{6}{*}{2} & 1100 & $\nicefrac{1}{2}(1+s_0)(1+s_1)(1+s_2)$ \\
& 1010 & $\nicefrac{1}{2}(1+s_0)(1+s_1)(1+s_3)$ \\
& 1001 & $\nicefrac{1}{2}(1+s_0)(1+s_1)(1+s_4)$ \\
& 0110 & $\nicefrac{1}{2}(1+s_0)(1+s_2)(1+s_3)$ \\
& 0101 & $\nicefrac{1}{2}(1+s_0)(1+s_2)(1+s_4)$ \\
& 0011 & $\nicefrac{1}{2}(1+s_0)(1+s_3)(1+s_4)$ \\
\hline

\multirow{4}{*}{3} & 1110 & $\nicefrac{1}{2}(1+s_0)(1+s_1)(1+s_2)(1+s_3)$ \\
& 1101 & $\nicefrac{1}{2}(1+s_0)(1+s_1)(1+s_2)(1+s_4)$ \\
& 1011 & $\nicefrac{1}{2}(1+s_0)(1+s_1)(1+s_3)(1+s_4)$ \\
& 0111 & $\nicefrac{1}{2}(1+s_0)(1+s_1)(1+s_3)(1+s_4)$ \\
\hline

4 & 1111 & $\nicefrac{1}{2}(1+s)(1+s_1)(1+s_2)(1+s_3)(1+s_4)$ \\	
\hline	
	\end{tabular}	
	\label{tbl:mfl}
\end{table}

\begin{table}[h]\small
\centering
	\caption{Path Fitness Landscape.}
	\begin{tabular}{c c c}
\hspace{0.2cm} Additional mutations \hspace{0.2cm} & \multirow{1}{*}{Phenotype} & \multirow{1}{*}{Birth probability} \\
\hline	
0 & 0000 & $\nicefrac{1}{2}(1+s)$ \\
\hline
\multirow{4}{*}{1} & 1000 & $\nicefrac{1}{2}(1+s_0)(1+s_1)$ \\
& 0100 & $\nicefrac{1}{2}(1 - v)$ \\
& 0010 & $\nicefrac{1}{2}(1 - v)$ \\
& 0001 & $\nicefrac{1}{2}(1 - v)$ \\
\hline

\multirow{6}{*}{2} & 1100 & $\nicefrac{1}{2}(1+s_0)(1+s_1)(1+s_2)$ \\
& 1010 & $\nicefrac{1}{2}(1 - v)$ \\
& 1001 & $\nicefrac{1}{2}(1 - v)$ \\
& 0110 & $\nicefrac{1}{2}(1 - v)$ \\
& 0101 & $\nicefrac{1}{2}(1 - v)$ \\
& 0011 & $\nicefrac{1}{2}(1 - v)$ \\
\hline

\multirow{4}{*}{3} & 1110 & $\nicefrac{1}{2}(1+s_0)(1+s_1)(1+s_2)(1+s_3)$ \\
& 1101 & $\nicefrac{1}{2}(1 - v)$ \\
& 1011 & $\nicefrac{1}{2}(1 - v)$ \\
& 0111 & $\nicefrac{1}{2}(1 - v)$ \\
\hline

4 & 1111 & $\nicefrac{1}{2}(1+s_0)(1+s_1)(1+s_2)(1+s_3)(1+s_4)$ \\	
\hline	
	\end{tabular}
	\label{tbl:pfl}
\end{table}

\smallskip\noindent{\em Density limit.}
Our tool allows a separate carrying capacity $K_i$ for each phenotype $i$.
When the \GFL\ is used, in the beginning of the simulation, the growth coefficients $s_i$ are calculated from the given $b_i$ for all phenotypes $i \in \{0,1 \}^4$ since the density limiting effects are based on the values for $\tilde{s_i}$.
As a technical detail, for sizes $X_i$ for which the birth probability $b_i$ would fall below 0 (or for which, equivalently, $\tilde{s_i}$ would fall below~-1), we set $b_i=0$.

\smallskip\noindent
\textbf{Statistics thread.}
The statistics thread handles the simulation of many identical branching processes to obtain the statistical results.
These simulations run in a separate thread such that the GUI keeps responsive for user requests.
After completing all the necessary simulations a data-file with all relevant results is automatically generated and stored to the execution directory of the tool.

\smallskip\noindent
\textbf{Graphical user interface.} 
The graphical user interface (GUI) component contains frames and forms required for the functionality of the tool. 
It also handles all the user requests and distributes them to the other components.
Within the GUI the plots for the tumor progression dynamics (in individual mode) and for the probability density of tumor detection (in statistics mode) are displayed. 
Multiple screenshots of the GUI are shown in section~\ref{screenshots}.

\begin{figure}[h]
	\centering
		\includegraphics[scale=0.5]{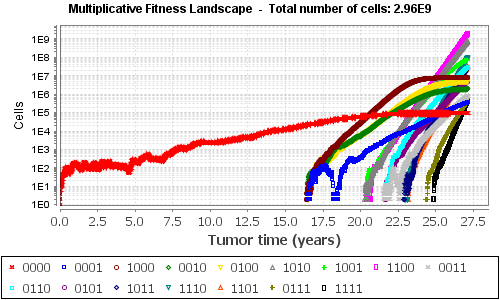}
	\caption{Example for a generated plot of the tumor growth dynamics.}
	\label{fig:ttp-plot-progression}
\end{figure}

\smallskip\noindent
\textbf{Plot generator.}
The plot generation is based on the free JFreeChart library (\texttt{\href{http://www.jfree.org}{www.jfree.org}}).
For the generation of the Scalable Vector Graphics (SVG) the Apache XML Graphics library (\texttt{\href{http://xmlgraphics.apache.org}{http://xmlgraphics.apache.org}}) is used. 
An example for a plot generated by our tool is shown in Figure \ref{fig:ttp-plot-progression}.

\smallskip\noindent
\textbf{Data files.} 
\tool\ produces various data files which can be used for further analysis and processing.
The data are given as tab-separated values where each record is one line of the text files.
In Listing \ref{lst:statfile} we show an example of a data file generated in the statistics mode. 
Average results are given as comments which start with a hash.
In the individual mode the data file contains the number of cells $X_i$ for each phenotype $i$ in all generations.

\begin{datafile}
\centering
\begin{lstlisting}[caption={Generated data file in the statistics mode.},
										label=lst:statfile,
										mathescape,language=Gnuplot,basicstyle=\scriptsize,tabsize=12]
# Used fitness landscape: Multiplicative
# Growth coefficients: s0=0.004, s1=0.006, s2=0.008, s3=0.01, s4=0.012
# Mutation rate: 1E-6
# Generation time (days): 3.0
# Simulation for 10000 patients.
# Bin size is 20 generations.
generation	detected	cumul_det	died	cumul_died
0	0	0	0.046003	0.920069
20	0	0	0.001938	0.958826
40	0	0	0.000705	0.972923
60	0	0	0.000358	0.98009
80	0	0	0.000227	0.984622
100	0	0	0.000149	0.987594
$\vdots$	$\vdots$	$\vdots$	$\vdots$ 	$\vdots$
5980	0	1	0	1
6000	0	1	0	1
# 0.7981% (abs: 10000) have been detected within at most 
# 	20000 generations.
# 99.2% (abs: 1243006) tumors went extinct.
# Average year of detection: 22.04
# Mutant appearance times: 
# Mutant 0000 appeared in average in generation: 0 (year: 0), 
# 	existance probability: 100%, 
# 	average number of cells: 3267013.74
# Mutant 0001 appeared in average in generation: 1540 (year: 12.66), 
# 	existance probability: 100%, 
# 	average number of cells: 1925660.02
$\vdots$	$\vdots$	$\vdots$	$\vdots$
# 1253006 runs have been performed to create 10000 patients.
\end{lstlisting}
\end{datafile}

\subsection{User Manual}
\begin{figure}[h]
	\centering
		\includegraphics[scale=0.75]{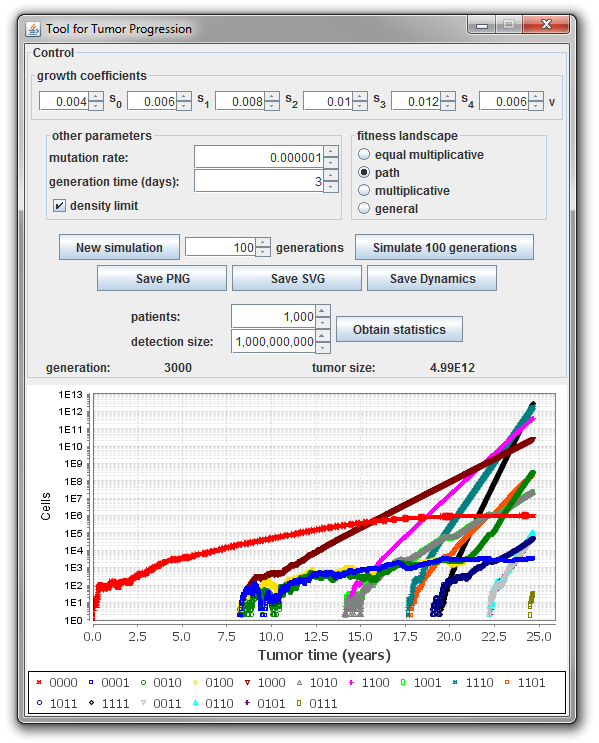}
	\caption{Graphical User Interface of TTP in the individual mode.}
	\label{fig:ttp-gui}
\end{figure}

\tool\ is invoked by a double-click on \texttt{ttp.jar} or by the command \texttt{java -jar ttp.jar}.
After the tool has started, the GUI can be used for all operations (see Figure \ref{fig:ttp-gui} for a screenshot of the GUI).

\smallskip\noindent
\textbf{Input parameters.} 
In the control panel the tool takes all the main parameters for tumor progression: the fitness landscape, the mutation rate, and the cell division time.
If one of the prespecified fitness landscapes (\MFL , \PFL, or \EMFL) is used, the relevant growth coefficients have to be defined.
If the general fitness landscape is used, a pop-up window appears after the selection of the \GFL\ and the specific birth probabilities for all the phenotypes can be defined.
To add density limits for specific phenotypes, a pop-up window appears after ``density limit'' has been checked. 
For each phenotype a different density limit can be given ($-1$ indicates that there is no limitation on this phenotype).
To obtain statistical results, the number of patients (i.e. the number of tumors with a surviving lineage) and the tumor detection size (i.e. the number of cells when a tumor can be detected) need to be provided.

\smallskip\noindent
\textbf{Modes.}
After all the parameter values have been specified, the tool can either run in the individual or the statistics mode.
To simulate the growth dynamics of a single tumor, click on ``New simulation'' and the tool runs in the individual mode.
Then any number of cell generations can be simulated until the tumor consists of more than $10^{15}$ cells.

The statistics mode can be started by clicking on ``Obtain statistics''.
The tool simulates the given number of tumors until they reach the detection size and calculates all relevant statistics.

\smallskip\noindent
\textbf{Output.}
In the individual mode, the tool generates plots for the growth dynamics and the phenotype distribution during the simulation.
Furthermore, the entire tumor growth dynamics for each phenotype can be stored as a data file and the plots can be saved as PNG and SVG files (plots are stored to the folder ``charts'' in the execution directory of the tool).

\begin{figure}[h]
	\centering
		\includegraphics[scale=0.75]{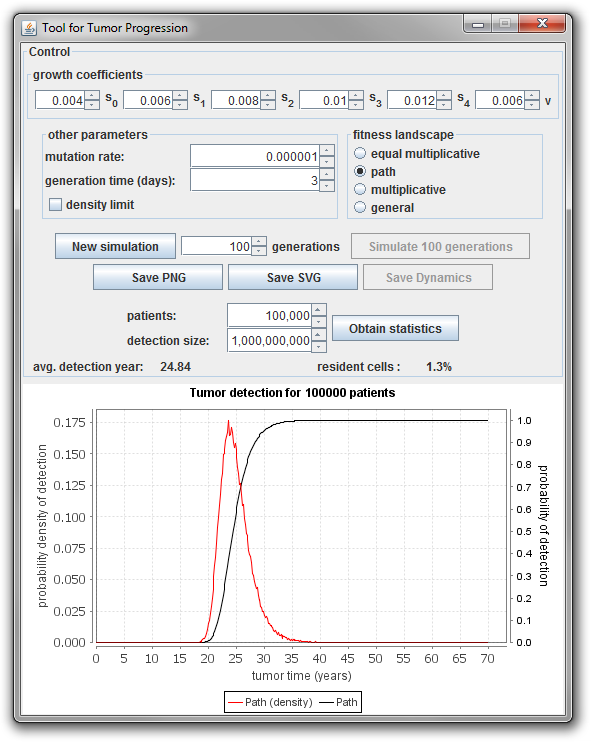}
	\caption{Graphical User Interface of TTP in the statistics mode.}
	\label{fig:ttp-guistat}
\end{figure}

\begin{figure}[b]
	\centering
		\includegraphics[scale=0.75]{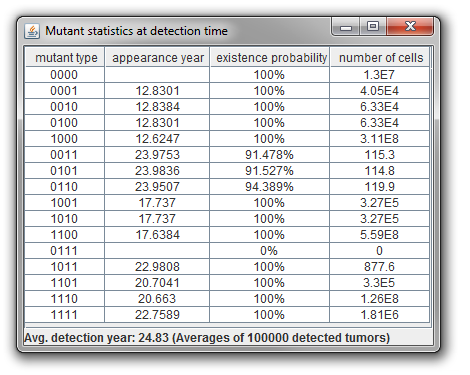}
	\caption{Statistical results of the average appearance year, the existence probability and the number of cells at detection time.}
	\label{fig:ttp-muts}
\end{figure}

In the statistics mode, the tool generates the plot for the probability density of tumor detection.
Statistics about the appearance time of the mutants and the detection and extinction year are shown in separate windows (see Figures \ref{fig:ttp-guistat} and \ref{fig:ttp-muts} for screenshots).
All the generated statistics are automatically saved to a data file (see Listing \ref{lst:statfile}).

\subsection{Experimental Results \& Screenshots} \label{screenshots}
In this section we present some additional experimental results and multiple screenshots of the tool.
In Table \ref{tbl:prob} we compare the probability of tumor detection for the fitness landscapes \EMFL , \MFL , and \PFL .
In average our tool needs approximately 90ms to simulate a tumor with $10^9$ cells (on a dual core 2.67GHz processor).

\begin{table}%
\centering
\caption{Cumulative probability of tumor detection for different fitness landscapes. Results are averages over $10^5$ runs. Parameter values: growth coefficients $s_0 = 0.004$, $s_1 = 0.006$, $s_2 = 0.008$, $s_3 = 0.01$, $s_4 = 0.012$, $v = 0.006$, mutation rate $u=10^{-6}$, detection size $M=10^9$.}
\begin{tabular}{c | c c c}
\hspace{0.3cm} generation \hspace{0.3cm} & \hspace{0.3cm} \EMFL \hspace{0.3cm} & \hspace{0.3cm} \MFL \hspace{0.3cm} & \hspace{0.3cm} \PFL \hspace{0.3cm} \\
\hline
2000		& 0.0		& 0.001		& 0.0 \\
2200		& 0.0		& 0.038		& 0.0 \\
2400		& 0.0		& 0.212		& 0.006 \\
2600		& 0.004		& 0.491		& 0.075 \\
2800		& 0.068		& 0.724		& 0.288 \\
3000		& 0.289		& 0.865 	& 0.559 \\
3200		& 0.567		& 0.937 	& 0.766 \\
3400		& 0.774		& 0.971 	& 0.887 \\
3600		& 0.892		& 0.987 	& 0.947 \\
3800		& 0.949		& 0.994 	& 0.976 \\
4000		& 0.977		& 0.998 	& 0.989 \\
4200		& 0.989		& 0.999 	& 0.995 \\
4400		& 0.995		& 1.0 	& 0.998 \\
4600		& 0.998		& 1.0 	& 0.999 \\
4800		& 0.999		& 1.0 	& 1.0 \\
5000		& 1.0		& 1.0 	& 1.0 \\
\end{tabular}
\label{tbl:prob}
\end{table}

%\begin{figure}[b]
%	\centering
%		\includegraphics[scale=0.75]{ttp-phenotypes.png}
%	\caption{Phenotype distribution of a tumor.}
%	\label{fig:ttp-phenotypes}
%\end{figure}

%\begin{comment}
%\end{comment}

\begin{figure}[h]
	\centering
		\includegraphics[scale=0.75]{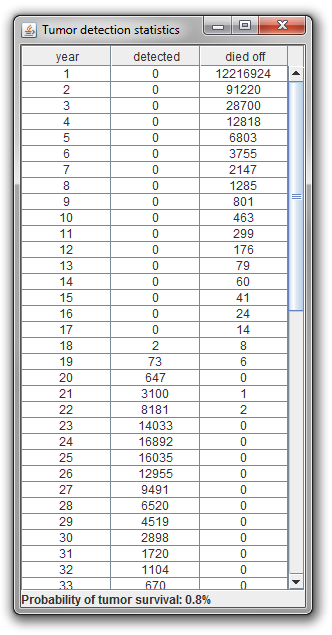}
	\caption{Statistical results of the average detection and extinction year.}
	\label{fig:ttp-stat}
\end{figure}

\end{document}